\documentclass[preprint]{aastex}

\begin{document}

\title{Chandra Observations of the X-ray Narrow-Line Region in \\ NGC 4151}

\author{Ogle, P. M., Marshall, H. L., Lee, J. C., and Canizares, C. R.}
\affil{MIT Center for Space Research}
\affil{NE80-6095, 77 Massachusetts Ave, Cambridge, MA 02139}
\email{pmo@space.mit.edu}

\shorttitle{Chandra Observations of NGC~4151}
\shortauthors{Ogle et al.}

\clearpage

\begin{abstract}

We present the first high-resolution X-ray spectrum of the Seyfert 1.5
galaxy NGC 4151. Observations with the Chandra High Energy Transmission 
Grating Spectrometer reveal a spectrum dominated by narrow emission lines 
from a spatially resolved (1.6 kpc), highly ionized nebula. The X-ray 
narrow-line region is composite, consisting of both photoionized and 
collisionally ionized components. The X-ray emission lines have similar 
velocities, widths, and spatial extent to the optical emission lines, showing 
that they arise in the same region. The clouds in the narrow-line region must 
contain a large range of ionization states in order to 
explain both the optical and X-ray photoionized emission. Chandra data give the
first direct evidence of X-ray line emission from a hot plasma 
($T\sim10^7$ K) which may provide pressure confinement for the cooler
($T=3\times10^4$ K) photoionized clouds. 
 
\end{abstract}


\section{Introduction}

NGC 4151 (z=0.00332) is a well-studied Seyfert 1.5 galaxy.
It is known to be extended in soft X-rays from Einstein and ROSAT observations
\citep{ebh83,mwe95}. The nature of the extended X-ray emission was postulated
to be either thermal emission from a collisionally ionized plasma or scattered
nuclear flux. \citet{wyh94} modeled the ASCA spectrum by 3 
components: an absorbed power law plus reprocessed emission from a warm 
absorber and an extra scattered component. Variability was observed in the 
column density of the warm absorber.

Spectral complexity from unresolved emission lines and ionized absorption 
edges has been seen in the residuals of power-law fits to the spectra of many 
Seyfert galaxies. Recent observations with Chandra have revealed
narrow absorption and emission lines in the spectra of Seyfert 1 galaxies 
NGC 5548 ~\citep{kml00} and NGC 3783 ~\citep{kbn00}. These two sources are 
dominated by continuum emission from the central source, so absorption lines 
are more prominent than emission lines. On the other hand, the nucleus of 
NGC 4151 is highly absorbed, giving a good view of its narrow emission line 
region. 

In this paper we present Chandra High Energy Transmission Grating 
Spectrometer (HETGS) observations of NGC 4151. We identify
and measure a large number of narrow emission lines in the  X-ray spectrum.
The high spectral resolution provided by HETGS allows us to study the 
ionization, temperature, and kinematics of the X-ray emitting regions. We 
discuss the association of the extended X-ray emission with the narrow-line 
region, and interpretations of the multiple X-ray emitting components in the 
spectrum.

\section{Observations}

We observed NGC 4151 for 48 ks on 2000 March 5 with Chandra  
HETGS ~\citep{c00}, at a roll angle of $153.2\arcdeg$. The dispersion 
directions of the High Energy Grating (HEG) and the Medium Energy Grating 
(MEG) made acute angles of $42.5\arcdeg$ and $37.5\arcdeg$, respectively, with
the direction of the extended X-ray emission. Since the source has appreciable
extent along the dispersion direction (see below),  the spectral and spatial 
dimensions are mixed. Spectral extractions and reductions were performed using
CIAO \footnote{http://asc.harvard.edu/ciao} and IDL. We used extraction 
windows of $7\farcs9$ width in the cross-dispersion direction, which were 
centered on the nucleus of the 0-order image. The windows included most of the
flux from the extended source. We then summed the plus and minus orders of the
spectra. The flux calibration of HETGS is currently accurate to 10\% 
for energies $E>2$ keV and 20\% for $E<2$ keV.

The wavelength calibration was tied to an HETGS observation of Capella which 
was reduced in an identical manner to NGC 4151 (using CIAO, v. 2.0$\alpha$, 
2000 August 6). In this way we determined that velocities must be corrected by
$-210\pm20$ km s$^{-1}$ to remove a systematic wavelength calibration error 
that exists in standard-processed  data. This error is currently attributed to 
an error in the CCD pixel scale, due to thermal contraction 
\footnote{http://asc.harvard.edu/cal}. Corrected velocities are reported in 
Table 1. Marginally redshifted X-ray narrow emission lines have been reported 
for NGC 5548 
~\citep[$270\pm100$ km s$^{-1}$]{kml00} and NGC 3783 
~\citep[$230\pm170$ km s$^{-1}$]{kbn00}. These redshifts may be owing to 
similar wavelength calibration errors.
 
NGC 4151 was in a low flux state during our observation, with 
$ 3\times10^{-3}$ photons s$^{-1}$ cm$^{-2}$ \AA$^{-1}$ at 2 \AA. We show the 
first order HEG and MEG spectra in Figure 1. Above 6 \AA~ (below 2.1 keV), the
spectrum is dominated by emission lines. Table 1 gives a list of the 
identified emission lines, their fluxes, and velocities relative to the galaxy
rest frame.

We compare the HETGS 0-order X-ray (0.4-2.5 keV) image of NGC 4151 to an 
optical [O {\sc iii}] 5007 image from the Hubble Space Telescope (HST) 
(Fig. 2). There is a good, but not exact correspondence between the structure 
of the optical and soft X-ray emission regions. The X-ray emission has a 
similar extent ($23\arcsec=1.6~h_{70}^{-1}$~kpc) and orientation 
(PA$=63\arcdeg$) to the optical narrow-line region (NLR). Most
(70\%) of the soft X-ray emission is resolved by Chandra; the rest comes from 
an FWHM$=0\farcs9$ nucleus. The hard X-ray emission (E$>$2.5 keV) is  
consistent with a nuclear point source.

The spatial profiles of the narrow emission lines are extended
(Figure 3). We show three lines, from three different ionization states:
O {\sc vii} (f), O {\sc viii} Ly$\alpha$, and Fe {\sc i} K$\alpha$. All three 
lines have similar spatial profiles to the 0-order soft X-ray image, and can 
be traced out to a distance of $3\arcsec=210~h_{70}^{-1}$~pc from the nucleus.

HETGS is rather insensitive to relativistically broadened Fe 
K$\alpha$ emission.  ~\cite{yew95} fit ASCA observations of Fe K$\alpha$ with 
broad ($\sigma=0.7$ keV) and narrow components having equivalent widths of 
$\sim 300$ eV and $\sim 100$ eV, respectively. We find an equivalent width of 
$160\pm20$ eV for narrow Fe {\sc i} K$\alpha$ with the HEG.  The narrow
core of the line is unresolved, with FWHM$=1800\pm 200$ km/s. 

\section{Discussion}
\subsection{Photoionized X-ray Line Emission}

The spectrum of NGC 4151 is hybrid, with line emission coming from both 
photoionized and collisionally ionized plasmas.  There are several strong
indications of a photoionized component. First, we see narrow radiative 
recombination continua (RRCs) from N {\sc vii}, O {\sc vii}, O {\sc viii}, 
Ne {\sc ix}, and Ne {\sc x}. We estimate half widths (at 37\% peak flux) for 
the RRCs which range from 1-4 eV, corresponding to temperatures of $2-4\times 
10^4$ K. To have such a low temperature and contain such high ionization 
states, the plasma where the RRC originates must be dominated by 
photoionization. 

The ratio of the O {\sc vii} forbidden to resonance emission lines 
($f/r=4.6\pm1.5$) is also consistent with photoionized plasma \citep{pd00}. In
addition, relatively strong $n=3-1$ transitions of N {\sc vi}, Ne {\sc ix},  
and Mg {\sc xi} are indicative of photoionization ~\citep{bk00}. The strength 
of  O, Ne, Mg, and Si K-shell emission lines and relative weakness of Fe 
L-shell lines are additional characteristics of a photoionized plasma 
 ~\citep{l99}. 

\subsection{Thermal X-ray Line Emission}

Except for O {\sc vii}, the He-like lines have a small $f/r$ ratio, which 
means that collisional ionization is important. The Ne {\sc ix} lines have 
$f/r=1.7\pm0.4$. A ratio $f/r\sim3.4$ is expected for a photoionized plasma, 
and $f/r<0.8$ for a collisionally ionized plasma with $T>2\times10^6$ K. There 
is clearly a mixture of line emission from both types of plasma. We note that 
fluorescent excitation by the continuum and resonance scattering may lead to 
enhancement of resonance lines under special circumstances 
 ~\citep{bka98,klo96}. However, we would expect these processes to equally 
effect the O {\sc vii} and other He-like resonance lines, which is not the 
case.

The Ly$\alpha$ lines of O {\sc viii}, Ne {\sc x}, and Mg {\sc xii} also have a
large contribution from collisionally ionized plasma. The O {\sc viii} 
Ly$\alpha$ line is much stronger than the O {\sc viii} RRC, while a ratio near
unity is expected for a pure photoionized plasma. We find that $89\pm5\%$ of 
the O {\sc viii} Ly$\alpha$ emission comes from the hot plasma component of 
the NLR. Similarly, $70\pm10\%$ of the Ne {\sc x} Ly$\alpha$ emission comes 
from this component. A large fraction of the Mg {\sc xii} Ly$\alpha$ flux must
also have a collisional origin, since the Mg {\sc xii} RRC (6.32 \AA) is not 
even detected. 

From the range of formation temperatures of the collisionally augmented lines,
we deduce that there is plasma in the temperature range $T=0.3-1\times10^7$ K. 
The nondetection of S {\sc xvi} emission (4.73 \AA, $2\pm5\times10^{-6}$ photons 
cm$^{-2}$ s$^{-1}$) is partly owing to the strong continuum at that 
wavelength. However, the detection of Fe {\sc xxv} ($T_f=8\times10^7$ K) 
suggests that hotter gas is present. A more thorough analysis will be 
necessary to derive the volume emission measure vs. temperature distribution 
of the hot plasma.

\subsection{Continuum Emission}

We characterize the continuum emission from NGC 4151 with the sum of two 
absorbed power laws. The hard continuum ($E>3$ keV) has a photon index 
$\Gamma=0.4\pm0.3$ and normalization 
$A = 2.3^{+1.6}_{-0.9}\times10^{-3}$ photons keV$^{-1}$ cm$^{-2}$ s$^{-1}$ at 
1 keV, absorbed by a neutral column of 
$N_H=3.7^{+1.1}_{-0.9}\times10^{22}$ cm$^{-2}$. The observed 2-10 keV flux is
$5.5 \times10^{-11}$ ergs cm$^{-2} s^{-1}$. The soft continuum ($E<1.3$ keV)
has a photon index of $\Gamma=3.1\pm0.5$ and normalization 
$A = 1.0\pm0.1\times10^{-3}$ photons keV$^{-1}$ cm$^{-2}$ s$^{-1}$ at 
1 keV, absorbed by a Galactic column of $N_H=2.0\times10^{20}$ cm$^{-2}$.

We deduce that there is strong absorption of the hard nuclear flux by a cloud
which does not occult the bulk of the soft X-ray emitting NLR.
\citet{wyh94} observed a 2-10 keV flux of $11-22\times10^{-11}$ ergs cm$^{-2} 
s^{-1}$, $\Gamma\sim1.5$, and $N_H=2-5\times10^{22}$ cm$^{-2}$  with ASCA.
The hard continuum was harder and a factor of $2-4$ times fainter during our 
observation, but $N_H$ was similar. However, we note that poor statistics for
energies above 7 keV in our data may bias the fit value for $\Gamma$.
\citet{wyh94} also found variations in $N_H$ with ASCA on a time scale of 6 
months, which suggests that absorption is taking place on a spatial scale 
much smaller than the NLR. 

The soft continuum of NGC 4151 has a much steeper photon index than the hard 
continuum (3.1 vs 0.4). This argues against Thomson scattered emission as the
source of the soft continuum, unless the intrinsic nuclear continuum steepens
drastically below 3 keV. Thermal bremsstrahlung and blended Fe L
emission lines are therefore the most likely sources of the soft continuum. 
 
\subsection{Kinematics of the X-ray Line Region}

The locations and velocity field of optically emitting clouds in the 
NLR of NGC 4151 have been mapped in detail by ~\cite{kbh00}, and are 
consistent with a biconical outflow from the nucleus. Integrated long-slit 
spectroscopy of the central $2\arcsec$ of the NLR gives a blueshift of -90 
km s$^{-1}$ and a line width (FWHM) of 460 km s$^{-1}$ for the [O {\sc iii}] 
5007 line \citep{hmb81}. The blueshifts and blue asymmetry of the optical 
lines are likely owing to obscuration of the far cone by dust. 

The unblended X-ray emission lines have an (unweighted) mean blueshift of
$-120\pm50$ km s$^{-1}$ with respect to the galaxy rest frame, as measured 
from its stellar spectrum.  This is consistent with the velocity of the 
optical NLR. Since the X-ray line widths are affected by spatial broadening, 
we can only get an upper limit to their velocity widths from the widths of the
low energy lines. O {\sc vii} (f) has an apparent width (Gaussian FWHM) of 
$460$ km s$^{-1}$, which equals the widths of the optical lines. 

It is interesting that the X-ray emission region is more extended in the front
(SW) cone than the back cone (Fig. 2), suggesting that the soft X-rays from 
the back cone are obscured, like the optical emission. The similar spatial 
extent of the X-ray and optical emission line regions (Figs. 2,3) and their 
similar kinematics are strong evidence that they are closely linked. X-ray 
emission from the NLR demonstrates that high energy processes are important in
this region, at distances of 70-800 pc from the galactic nucleus.

\subsection{The Multiphase Nature of the NLR}

For the first time, there is direct evidence for at least 2 distinct gas phases
in the extended NLR of a Seyfert galaxy. One of the phases
is cool ($T=3\times10^4$ K) and photoionized, while the second is hot 
($T\sim10^7$ K) and collisionally ionized. In addition, there is a
large range in ionization of the photoionized region, as demonstrated by
O {\sc i}-O {\sc iii} in the optical-UV spectrum and O {\sc vii} in the X-ray 
spectrum. There are also strong K$\alpha$ lines from neutral species in the 
X-ray spectrum, including Fe {\sc i}, Si {\sc i}, and Mg {\sc i}.  
Detailed modeling will be necessary to determine the volume emission measure 
vs. ionization distribution of the photoionized component. A wide range of 
ionization is derived from photoionization models of the NLR of the Circinus 
galaxy ~\citep{skp00}, and might be explained by a radial gradient in the NLR 
density.  

A large percentage ($65\pm9\%$) of the narrow Fe {\sc i} K$\alpha$ emission 
comes from the extended NLR (Fig. 3). This is contrary to the popular 
idea that the narrow emission comes primarily from a parsec scale torus. 
The remaining unresolved emission, which has an equivalent width of 
$56\pm9$ eV, may very well come from a torus. Unlike NGC 5548 \citep{ygn00}, 
there is no evidence for Doppler broadening of the narrow Fe {\sc i} K$\alpha$ 
line in NGC 4151. 

It has been suggested that a hot, inter-cloud medium is necessary for
pressure confinement of the NLR clouds ~\citep{ebh83}. The hot plasma 
we detect in our X-ray spectrum may form this medium. We derive volume 
emission measures for the thermal contribution to the Ne {\sc x} and O 
{\sc viii} Ly$\alpha$ lines of 
$n_e^2V=2-4\times10^{63} h_{70}^{-2}$ cm$^{-3}$, assuming 
$T=1\times10^7 K$, and line emissivities for solar abundances from 
\cite{mgv85}. ~\cite{ckh00} derive a half-opening angle of $36\arcdeg$ from 
kinematic models of the optical NLR. If the thermal X-ray line emission comes 
from a uniformly filled ($f=1$) bicone in the central $4\arcsec$ of the 
galaxy, then the mean plasma density is $n_e=3$ cm$^{-3}$ and the pressure is 
$n_eT=3\times10^7$ cm$^{-3}$ K. Pressure equilibrium between the hot and cool 
phases of the X-ray NLR would then require a density of $n_e=10^3$ cm$^{-3}$ 
for the cool phase.

The optical NLR clouds at $r=6-20\arcsec$ have a density $n_e=220$ cm$^{-3}$
and a pressure of $n_eT=3\times10^6$ cm$^{-3}$ K~\citep{pra90,mwe95}, much
lower than what we find closer to the nucleus. However, we know 
that the spatial profile of the NLR flux is centrally peaked. A density 
profile of $r^{-2}$ would account for the larger pressure of the nucleus than 
the extended NLR, assuming constant temperature. Estimates of the pressure 
can be improved by modeling the density distribution to match the spatial
profiles of the X-ray emission.

A possible alternative is that the hot plasma is confined to shocks between 
the NL clouds and the host galaxy ISM. In that case, its filling factor
$f$ would be much lower, and its density and pressure much higher than 
calculated above. The X-ray emission we observe in the Chandra image has a 
clumpy morphology which follows the optically emitting clouds. This may be 
partly owing to emission from a shocked component. There is evidence for 
deceleration of the NL clouds at a radius of about $2\arcsec$ from the nucleus
 ~\citep{ckh00}, perhaps from interaction with the ISM. 

\section{Conclusions}

The power of Chandra HETGS to elucidate the nature of the X-ray emission
from active galactic nuclei is demonstrated by high resolution spectra
and images of NGC 4151. Direct emission from the nucleus is highly absorbed. 
Strong, narrow X-ray emission lines are seen from the spatially resolved NLR 
of NGC 4151. The narrow radiative recombination continuum features and the 
O {\sc vii} $f/r$ ratio are consistent with photoionization from the hard 
nuclear X-ray source. There is also narrow line emission from a hot, 
collisionally ionized component in the extended NLR. This may come from the 
long-sought intercloud medium or shocked plasma. In addition, we find strong
narrow Fe {\sc i} K$\alpha$ emission from the extended NLR. The composite 
nature of the NLR indicates that both photoionization and collisional heating 
are important. The extent and kinematics of both X-ray photoionized clouds and
hot thermal plasma match those of the optical NLR, showing that they are 
closely connected. We also suggest that the narrow X-ray emission lines seen 
in some other Seyfert galaxies may come from the NLR, and may be interpreted 
in the light of these NGC 4151 observations.

\acknowledgements

We thank everyone whose hard work made Chandra and HETGS possible. We also
thank the referee, Kim Weaver, for many helpful comments.
This work was funded in part by contracts NAS8-38249 and SAO 
SV1-61010. Some of the data presented in this paper were obtained from the 
Multimission Archive at the Space Telescope Science Institute (MAST). STScI 
is operated by the Association of Universities for Research in Astronomy, 
Inc., under NASA contract NAS5-26555.

\clearpage

\begin{deluxetable}{lcccl}
\tablecolumns{5}
\tablewidth{10cm}
\tabletypesize{\small}
\tablecaption{NGC 4151 Emission Lines}
\tablehead{\colhead{$\lambda$(rest \AA)} &\colhead{v(km s$^{-1}$)}& 
           \colhead{Flux}\tablenotemark{a} 
           &\colhead{SNR}\tablenotemark{b} &\colhead{ID}}

\startdata
1.79   & 460$\pm$420 & 3.3E-05  & 2.3  & Fe {\sc xxv} \\
1.937  & 250$\pm$80  & 1.8E-04  &11.0  & Fe {\sc i} K$\alpha$ \\
6.182  & -20$\pm$120 & 1.0E-05  & 4.1  & Si {\sc xiv} Ly$\alpha$ \\
6.648  & -30$\pm$130 & 1.1E-05  & 4.4  & Si {\sc xiii} r \\
6.740  & -40$\pm$40  & 1.1E-05  & 5.0  & Si {\sc xiii} f \\
7.106  &-380$\pm$110 & 5.3E-06  & 3.5  & Mg {\sc xii} Ly$\beta$ \tablenotemark{c} \\
7.130  &-380$\pm$40  & 1.3E-05  & 5.6  & Si {\sc i} K$\alpha$   \\
7.851  &-560$\pm$80  & 2.5E-06  & 2.4  & Mg {\sc xi} 1s3p-1s$^2$\\
8.421  &-210$\pm$90  & 1.1E-05  & 7.4  & Mg {\sc xii} Ly$\alpha$ \\
9.01   &-380$\pm$170 & 3.5E-06  & 3.3  & Fe {\sc xxii} \tablenotemark{c}\\
9.102  &\nodata      & 7.2E-06  & 3.2  & Ne {\sc x} RRC \tablenotemark{c} \\
9.169  & -20$\pm$160 & 9.3E-06  & 6.3  & Mg {\sc xi} r \tablenotemark{c}\\
9.314  &-406$\pm$140 & 1.2E-05  & 7.4  & Mg {\sc xi} f  \\
9.887  &-330$\pm$50  & 3.2E-06  & 2.7  & Mg {\sc i} K$\alpha$  \\
10.239 &-180$\pm$70  & 7.4E-06  & 4.5  & Ne {\sc x} Ly$\beta$ \\
10.368 &\nodata      & 9.3E-06  & 4.2  & Ne {\sc ix} RRC \\
11.001 & 110$\pm$140 & 3.8E-06  & 2.2  & Ne {\sc ix} 1s4p-1s$^2$\\
11.547 &-290$\pm$70  & 5.9E-06  & 3.0  & Ne {\sc ix} 1s3p-1s$^2$\\
12.134 & -70$\pm$80  & 2.0E-05  & 6.4  & Ne {\sc x} Ly$\alpha$ \\
13.447 &-150$\pm$40  & 2.0E-05  & 4.8  & Ne {\sc ix} r \\
13.553 &-210$\pm$160 & 9.9E-06  & 2.5  & Ne {\sc ix} i \\
13.698 &  10$\pm$40  & 3.3E-05  & 6.1  & Ne {\sc ix} f \\
14.228 &\nodata      & 1.1E-05  & 2.3  & O {\sc viii} RRC \tablenotemark{c}\\
15.176 &-170$\pm$50  & 1.1E-05  & 3.2  & O {\sc viii} Ly$\gamma$ \\
16.006 &-250$\pm$20  & 1.8E-05  & 3.8  & O {\sc viii} Ly$\beta$ \tablenotemark{c} \\
16.771 &\nodata      & 5.2E-05  & 4.7  & O {\sc vii} RRC \\
18.588 &\nodata      & 2.6E-05  & 2.6  & N {\sc vii} RRC \\
18.969 & -40$\pm$30  & 1.0E-04  & 7.0  & O {\sc viii} Ly$\alpha$ \\
21.602 & 106$\pm$20  & 6.8E-05  & 4.1  & O {\sc vii} r \\
21.804 &-410$\pm$140 & 5.1E-05  & 3.3  & O {\sc vii} i \\
22.101 &-132$\pm$20  & 3.1E-04  & 8.3  & O {\sc vii} f \\
24.781 &-190$\pm$60  & 6.3E-05  & 4.0  & N {\sc vii} Ly$\alpha$\\
24.898 & -10$\pm$20  & 3.9E-05  & 3.0  & N {\sc vi} 1s3p-1s$^2$\\
\enddata

\tablenotetext{a}{Flux (photons s$^{-1}$ cm$^{-2}$)}
\tablenotetext{b}{Signal-to-noise ratio}
\tablenotetext{c}{Line blend or possible line blend}
\end{deluxetable}

\clearpage
\centerline{\bf Figure Captions}

\noindent
{\bf Figure 1}\\
Chandra first order HEG (top) and MEG (middle, bottom)
spectra of NGC 4151, with line identifications. Above 6 \AA, the spectrum is 
dominated by narrow emission lines from both thermal and photoionized plasmas.
Note that the ratio of forbidden to resonance emission in the He-like O 
{\sc vii} lines is large, while the ratio for Ne {\sc ix}, Mg {\sc xi}, and Si
{\sc xiii} is smaller, indicating a hybrid spectrum. The narrow radiative 
recombination continua (RRCs) of several ions attest to photoionization in a 
cool $T=3\times10^4 K$ plasma.

\noindent
{\bf Figure 2}\\
Comparison of X-ray and optical emission from the extended
NLR in NGC 4151.  Chandra contours (0.4-2.5 keV band) are overlayed on an 
HST [O {\sc iii}] 5007 image. A 36$\farcs$4 by 23$\farcs$5 region is shown,
with N at top and E at left. Note that the SW (front) cone is more extended 
than the NE (back) cone in both bands. X-ray contours are separated by
factors of 2 in surface brightness, with the lowest contour at 2.1 photons
arcsec$^{-2}$.

\noindent
{\bf Figure 3}\\
Comparison of NGC 4151 zero-order spatial profile and 1st order 
emission line spatial profiles. The gaussian profile of the unresolved
nuclear component is shown for comparison. The spatial profiles of 
O {\sc vii}, O {\sc viii}, and Fe K$\alpha$ are all resolved, from a region at
least $6\arcsec$ in extent.

\begin{figure}
\plotone{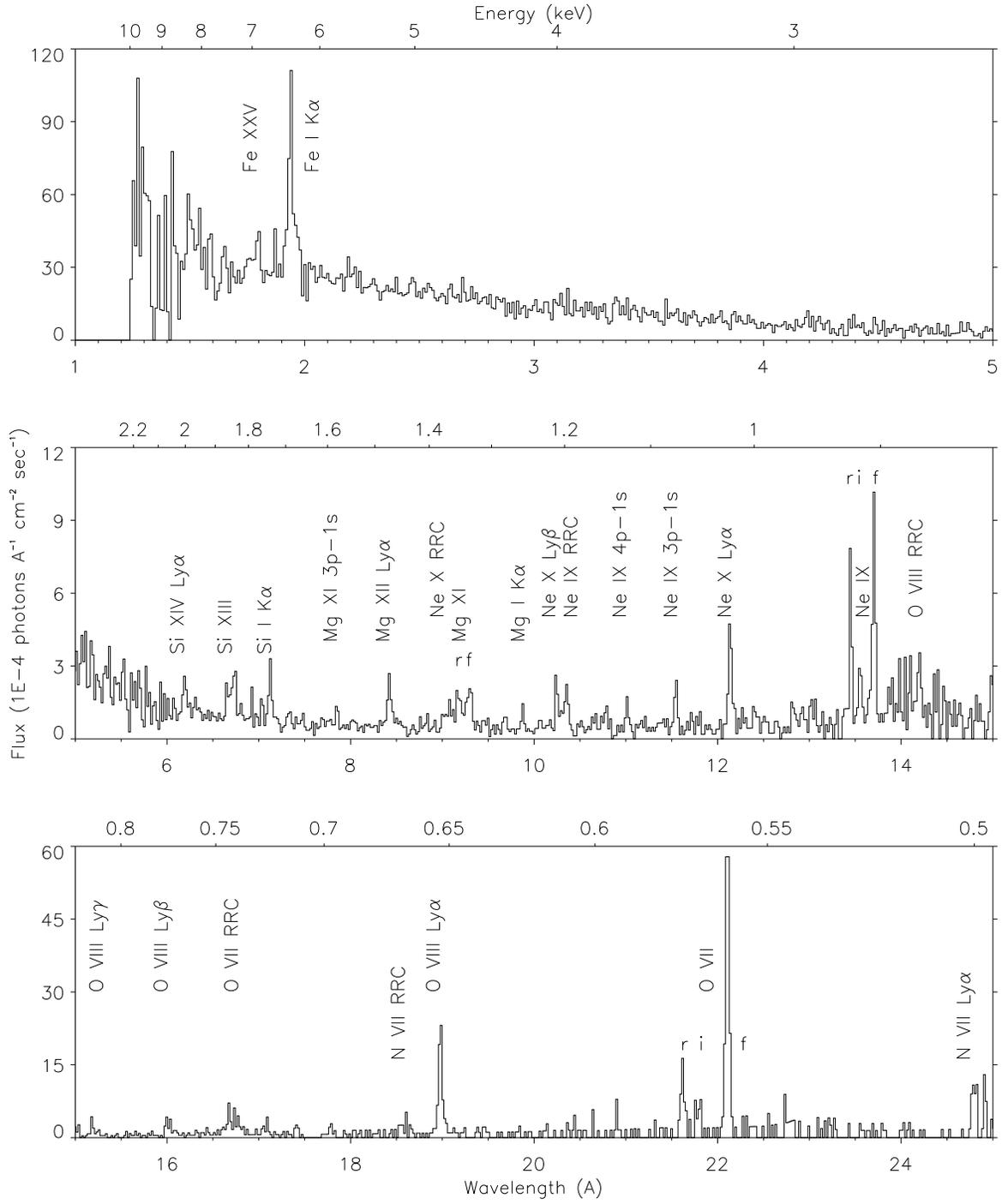}
\figcaption{See figure caption.}
\end{figure}

\begin{figure}
\plotone{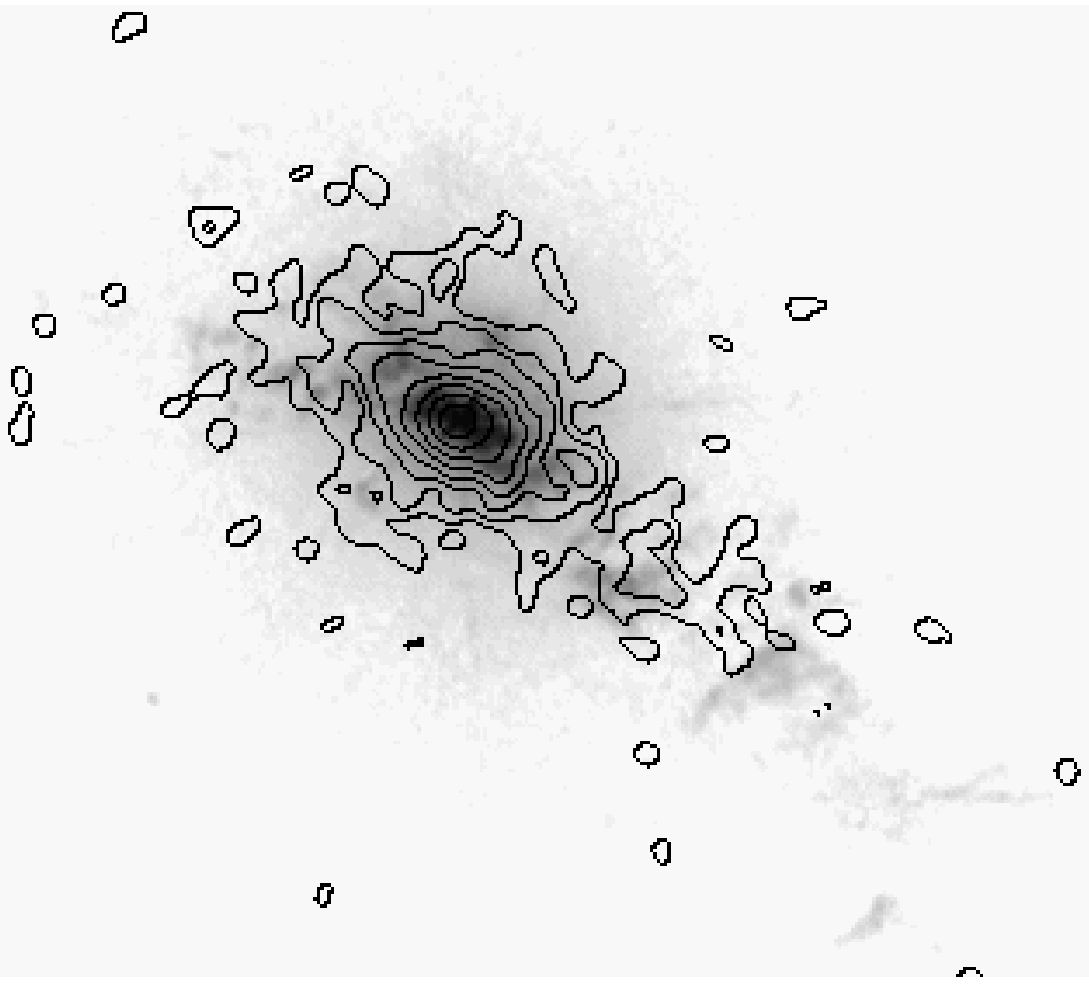}
\figcaption{See figure caption.}
\end{figure}

\begin{figure}
\plotone{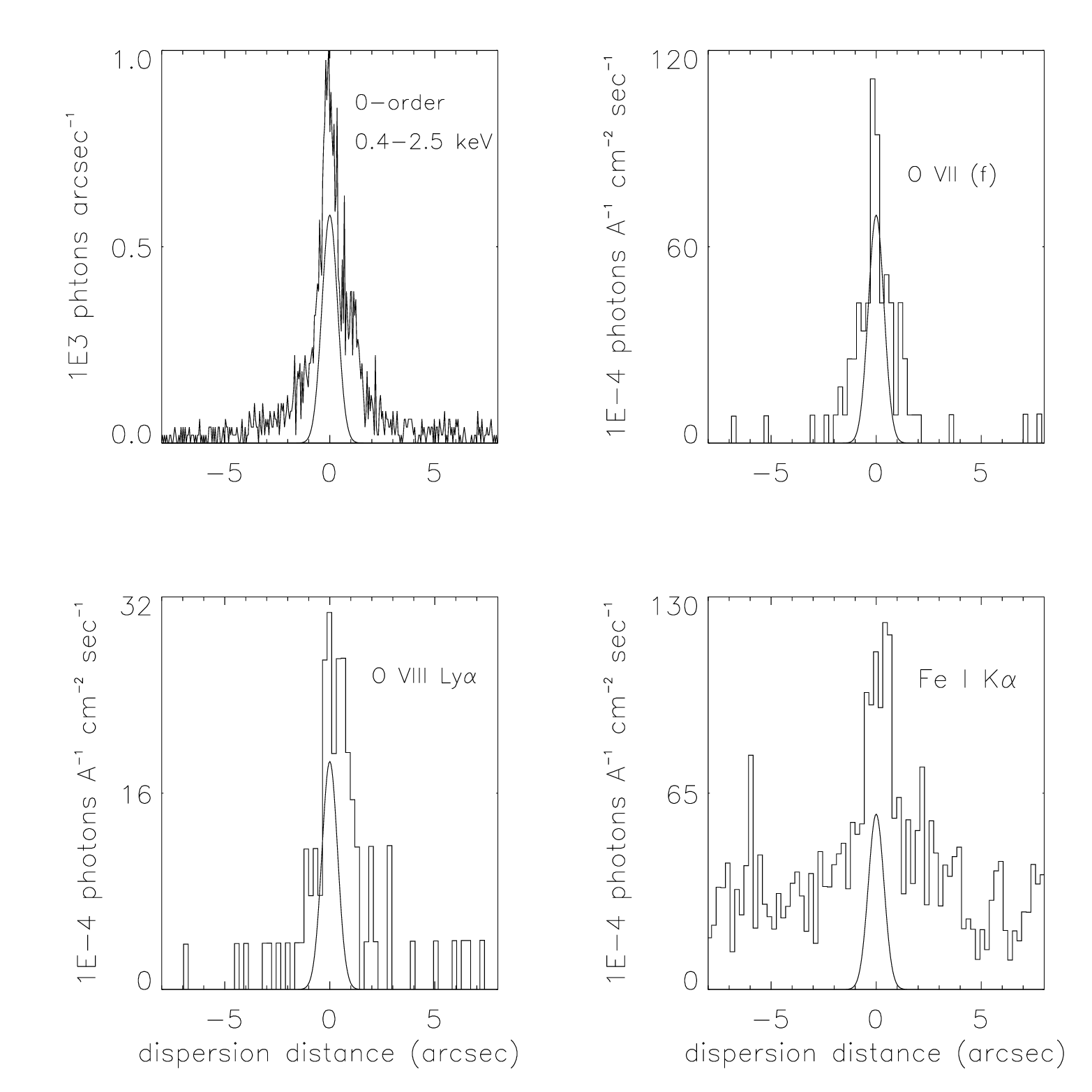}
\figcaption{See figure caption.}
\end{figure}

\end{document}